# Edge superconductivity in Multilayer WTe2 Josephson junction

Ce Huang[1,2†], Awadhesh Narayan[3†], Enze Zhang[1,2], Xiaoyi Xie[1,2], Linfeng Ai[1,2], Shanshan Liu[1,2], Changjiang Yi[4], Youguo Shi[4,5], Stefano Sanvito[6], Faxian Xiu[1,2,7*]

[1] State Key Laboratory of Surface Physics and Department of Physics, Fudan University, Shanghai 200433, China

[2] Collaborative Innovation Center of Advanced Microstructures, Nanjing 210093, China

[3] SSCU, Indian Institute of Science, Bengaluru 560012, India

[4] Institute of Physics and Beijing National Laboratory for Condensed Matter Physics, Chinese Academy of Sciences, Beijing 100190, China

[5] School of Physical Sciences, University of Chinese Academy of Sciences, Beijing 100190, China

[6] School of Physics, AMBER and CRANN Institute, Trinity College, Dublin 2, Ireland

[7] Institute for Nanoelectronic Devices and Quantum Computing, Fudan University, Shanghai 200433, China

[†]These authors contributed equally to this work.

[*]Correspondence and requests for materials should be addressed to F. X. (E-mail: Faxian@fudan.edu.cn).

## Abstract

WTe2, as a type-II Weyl semimetal, has 2D Fermi arcs on the (001) surface in the bulk and 1D helical edge states in its monolayer. These features have recently attracted wide attention in condensed matter physics. However, in the intermediate regime between the bulk and monolayer, the edge states have not been resolved owing to its closed band gap which makes the bulk states dominant. Here, we report the signatures of the edge superconductivity by superconducting quantum interference measurements in multilayer WTe2 Josephson junctions and we directly map the localized supercurrent. In thick WTe2 ($\sim 60\,\mathrm{nm}$), the supercurrent is uniformly distributed by bulk states with symmetric Josephson effect ($|I_c^+(B)| = |I_c^-(B)|$). In thin WTe2 (10 nm), however, the supercurrent becomes confined to the edge and its width reaches up to $1.4\,\mathrm{\mu m}$ and exhibits non-symmetric behavior $|I_c^+(B)| \neq |I_c^-(B)|$. The ability to tune the edge domination by changing thickness and the edge superconductivity establishes WTe2 as a promising topological system with exotic quantum phases and a rich physics.

## Introduction

Layered WTe$_2$ was suggested as the first material candidate to be a type-II Weyl semimetal, where eight separated Weyl points exist in the bulk and topological Fermi arcs occur on the (001) crystal surfaces owing to the reflection symmetry[1]. An extra set of quantum oscillations arising from Weyl orbits were observed as evidence of Fermi arcs in transport[2]. Intriguingly, when the thickness is reduced to the monolayer, WTe$_2$ turns to be a quantum spin Hall insulator with edge states[3], which have been demonstrated in numerous experiments involving low-temperature transport[4, 5], angle-resolved photoelectron spectroscopy[6], scanning tunneling microscopy[7, 8] and microwave impedance microscopy[9]. Besides, it has also been predicted that WTe$_2$ has 1D hinge states as a higher-order topological insulator[10].

While the boundary modes of WTe$_2$ have been well studied in both the 3D and 2D limits[11, 12], in multilayers these modes become rather complicated due to the intervening bulk and edge states and thus they remain largely unexplored. Unlike the monolayer WTe$_2$, the nearly-closed bandgap in multilayer WTe$_2$ results in a large density of bulk states. Therefore, it is difficult to distinguish the edge states through a gating approach. It is then necessary to make them distinct from the coexisting bulk ones. However, separating edge and bulk states in a single electrical conductance measurement may be ambiguous. In contrast, if the charge carriers condense together to form Cooper pairs, the difference can be amplified since the supercurrent properties are largely related to the coherence length[13]. A planar microscopic Josephson junction to realize superconducting TSM is feasible to elucidate the boundary states. For example, in Nb/Bi$_2$Te$_3$/Nb Josephson junctions the surface states enable the ballistic Josephson current rather than the diffusive bulk transport[14]. The supercurrent distribution in real space can be also quantitatively extracted from the superconducting quantum interference (SQI) measurements, where a perpendicular magnetic field induces oscillations in the amplitude of the superconducting current in Josephson junctions. This approach has been widely adopted to reveal the quantum spin Hall edge states in HgTe quantum well[15], topological surface states in TI[16] and quantum Hall edge states in graphene[17], but not yet in TSM.

Here, we report the observation of edge superconductivity in multilayer WTe$_2$ Josephson junctions. By varying the thickness of WTe$_2$ in SQI experiments, we have observed the Fraunhofer and the mixture of Fraunhofer and SQUID pattern in thick and thin WTe$_2$, respectively, which indicates the edge superconductivity in thin WTe$_2$. The non-uniform supercurrent exists in multilayer WTe$_2$ up to 16 nm, while the bulk supercurrent density amplitude ($J_c$) is about $1/3$ of the edge in the thinnest sample. $|I_c^+(B)| \neq |I_c^-(B)|$ is also observed in thin WTe$_2$ due to the inversion symmetry breaking.

## WTe$_2$ Josephson junctions

We measure several Josephson junctions consisting of WTe$_2$ flakes of different thicknesses contacted by niobium (Nb) leads. The fabrication and characterization details are described in the Methods and SI Section I (Supplementary Fig. 1-2). A scanning electron microscopy (SEM) image of the actual device and its measurement configuration are displayed in Fig. 1a (device #1, 10 nm-thick WTe$_2$). The length and width of the superconducting channel are $L = 200$ nm and $W = 13$ μm, respectively. Figure 1b shows the resistance-temperature ($R$-$T$) curve of the junction with two transitions $T_{c1}$ and $T_{c2}$ at zero magnetic field. $T_{c1} \sim 8$ K originates from the Nb superconducting transition, while the resistance continues to drop to $10^{-3}$ times of the normal resistance below $T_{c2} \sim 0.72$ K which comes from the proximity Josephson coupling. The Josephson effect is highly reproducible across different devices, as shown in Supplementary Fig. 3. Figure 1c and its inset display the $I$-$V$ characteristics and the differential resistance ($dV/dI$) of the junction at 45 mK, respectively. From the slope of the $I$-$V$ curve in the high bias region ($I > 10$ μA) where the curve is linear, the normal-state resistance $R_N \sim 1.7$ Ω is extracted. For $|I| < 4.1$ μA, the voltage and $dV/dI$ across the junction remain nearly zero, indicating a robust Josephson effect. Ten WTe$_2$ Josephson junctions with various $L$ and $W$ are studied (see Table S1 for their junction parameters), all exhibiting a finite supercurrent at low temperatures with reproducible behavior. The junction is in the long junction limit[18] (see SI Section IV and Supplementary Fig. 4 for details). Therefore it follows the $1/L$ dependence found from the $I_c R_N$ plot versus $L$ in Fig. 1d. In this long junction regime, the critical current is given[18, 19] by $I_c \sim \frac{E_{Th}}{e R_N}$, being determined by the Thouless energy $E_{Th}$, which can be estimated[20] to be $\sim \hbar v_F / L$, yielding $I_c R_N \propto 1/L$

**The superconducting quantum interference measurements**

Having established the Josephson effect in our Nb/WTe$_2$/Nb junctions, we then focus on the supercurrent of WTe$_2$. In our experiments, the spatial distribution of the supercurrent is analyzed by applying a perpendicular magnetic field $B$ during the SQI measurements with different thicknesses of WTe$_2$. The particular shape of the critical current interference pattern depends on the phase-sensitive summation of the supercurrents traversing the junction. In the case of a symmetric supercurrent distribution, this integral takes the simple form[21]:

$$I_c^{max}(B) = \left| \int_{-\infty}^{\infty} J_c(x) \cos\left(\frac{2\pi L_{eff} B x}{\Phi_0}\right) dx \right|$$

where $L_{eff}$ is the effective length of the junction along the direction of the current, accounting for the magnetic flux threading[22] through parts of the superconducting contacts over the London penetration depths. $\Phi_0 = h/2e$ corresponds to the magnetic flux quantum.

In thick WTe$_2$, the bulk states dominate and along the $y$-axis the supercurrent density has an approximately uniform distribution as shown in Fig. 2a. Thus, the uniform current density yields the single-slit Fraunhofer pattern described by[21]

$$I_c(B) = I_{c0} \left| \sin\left(\frac{\pi L_{eff} B W}{\Phi_0}\right) \middle/ \left(\frac{\pi L_{eff} B W}{\Phi_0}\right) \right|.$$

We have measured device #8 with 60 nm-thick WTe$_2$ as shown in Fig. 2b. The critical current envelope has an oscillation characteristic. We obtain a period of $\sim 0.33$ mT, which yields the effective length of $L_{eff} = \Phi_0/(\delta B_{lobe} W) \sim 1.0$ μm. This effective length, larger than the distance between the two Nb electrodes ($L = 240$ nm), is caused by the London penetration depth and the flux focusing due to the Meissner effect[23, 24]. The critical current envelope strongly resembles a single-slit pattern with $2\Phi_0$ central lobe width. The corresponding supercurrent distribution is obtained by transforming the single-slit pattern to the real-space current density, $J_c(z)$, as shown in Fig. 2c. This suggests a nearly uniform supercurrent density throughout the $y$ direction. The full details of the extraction procedure can be found in the SI section V and Supplementary Fig. 5. Furthermore, the critical currents overlap each other at different current directions as shown in Fig. 2d which indicates a symmetric Josephson effect with $|I_c^+(B)| = |I_c^-(B)|$, where $+$ and $-$ denote the sweep direction of the bias current and $+B$ and $-B$ are the magnetic field directions.

When the WTe₂ is thinned down to a few layers, the low bulk density of states coexists with the possible high density at edges as shown in Fig. 2e. The magnetic-field-dependent critical current envelope in a 13 nm-thick WTe₂ device (device #2) demonstrates the periodic SQI with a $1.6\Phi_0$ central lobe width (Fig. 2f). $I_c$ decays slowly which is distinct from the Fraunhofer pattern in Fig. 2b. We use an edge-stepped nonuniform supercurrent model to directly simulate the $I_c - B$ relation as shown by the black line in Fig. 2f (see the model details in SI Section VI and Supplementary Fig. 6). The good fit of both the magnitude and periodicity of $I_c$ indicates the nonuniform supercurrent in few-layer WTe₂. Furthermore, the mixture of Fraunhofer and SQUID interference pattern corresponds to the development of sharp peaks in the supercurrent density at the mesa edges in Fig. 2g. The widths of the supercurrent-carrying edge are estimated to be in the range $1.3 - 1.4 \, \mu m$. The value is similar to other edge superconductivity systems[15, 25] and the additional edge modes or bulk modes coupled weakly to edge states across the junction to carry supercurrent can result in the large edge supercurrent channel. The relation of the critical current with the magnetic field is presented in Fig. 2h and behaves non-symmetric $|I_c^+(B)| \neq |I_c^-(B)|$ in most magnetic fields which is different with thick sample.

We have reproduced the edge superconductivity and the mixture of Fraunhofer and SQUID pattern in 10 nm-thick WTe₂ device (#1), and the traditional Fraunhofer pattern in a 40 nm-thick WTe₂ device (#5, see SI Section VII and Supplementary Fig. 7 for details). The higher supercurrent density at edges suggests a robust coupling to the superconductor electrodes.

To further distinguish the superconducting proximity Josephson coupling of edge/bulk, we experiment with a 16 nm-thick WTe₂ device (#3) to distinguish the bulk and edge contributions. Two Josephson channels are fabricated as the edge-crossing ($R_1$) and edge-untouched ($R_2$) as shown in Fig. 3a. For $R_2$, the junction is easier to be conducted by the bulk because the electrodes are closer in the central region ($L_b \sim 0.4 \, \mu m$) while far at the edge. The distance on the edge side is $L_s \sim 4 \, \mu m$ that makes it hard to realize the Josephson effect through the edge region. Since the thickness is uniform in this sample, as indicated by the atomic force microscopy (AFM) measurement (Supplementary Fig. 2c-d), we can reasonably assume that the resistance by bulk states is isotropic and inversely proportional to the width. Figure 3b shows the $R$-$T$ curve at low temperatures. Only edge-crossing $R_1$ can reach zero to exhibit Josephson effect while $R_2$ only decreases a little. The differential resistance versus the current measurement in Fig. 3c verifies this property. Since the lengths of two junctions differ slightly, the coherence length of $R_1$ should be larger than that of $R_2$ to realize the Josephson effect. A similar mixture of Fraunhofer and SQUID pattern with edge-dominated supercurrent is also observed, as shown in Supplementary Fig. 7b, which is consistent with the other two thin devices (#1 and #2). On the contrary, $R_2$ does not exhibit any oscillation and only the central lobe is observed (see Supplementary Fig. 8 for details). The width $W_2$ for $R_2$ is estimated to be 1.9 μm, which corresponds well to the actual junction width 1.7 μm as shown in Fig. 3a. The in-complete superconductivity of $R_2$ is due to the weaker superconducting combining for bulk. If the Josephson channel is further shortened, the bulk-only channel $R_2$ can also be superconducting in another device #9. However, only the Fraunhofer pattern with uniform supercurrent density is observed and corresponds well to the bulk-dominated sample #8 in Fig. 2c-d (see Figs. S8-9 for details).

**Discussion**

It is necessary to discuss whether the observed edge superconductivity originates from the edge states in $WTe_2$ or other trivial effects. All of the four different devices exhibit the sharp edge superconductivity which can exclude the accidental impurity effect. The exclusion of some trivial effects such as fluctuations and the affection by the $SiO_2$ substrate and the capping layer in thinner $WTe_2$ is also discussed in Supplementary Section VIII. However, it is difficult to exclude other trivial effects such as trivial edge states. Moreover, the other trivial mechanisms can also lead to a similar non-uniform supercurrent such as an inhomogeneous interface. A mixture of Fraunhofer and SQUID pattern was also observed in Nb-InGaAs/InP junctions with a step-shaped current density distribution[26]. Therefore, we need to point out that the edge superconductivity we observed is not equivalent to the superconductivity in the edge modes nor any evidence of toplogical superconducting phase. On the contrary, only the superconductivity in the edge region of samples can be concluded in our experiments.

The critical currents following $|I_c^+(B)| \neq |I_c^-(B)|$ in thin $WTe_2$ is quite interesting. In general, the asymmetric crystal can induce different Fermi velocities at two sides and results in supercurrent asymmetry. Since the supercurrent density is uniform as shown in Fig. 2c with symmetric Fraunhofer pattern (Fig. 2d), the bulk $WTe_2$ does not contribute to the asymmetry. Consequently, this supercurrent asymmetry may be related to the edge which is consistent with the predicted effect of inversion-symmetry-breaking on Weyl semimetal[27]. The total Josephson current carried by the two edges can be described by[28]

$$I(\Phi, \varphi) \propto I_1 \sin(n\varphi + n\Phi) + I_2 \sin(n\varphi - n\Phi)$$

where $I_1$ and $I_2$ represents the Josephson current carried by the two edges, $\Phi$ and $\varphi$ are the phase in $WTe_2$ ( the magnetic-field-related) and Nb regions (the current-related), respectively. The two edges have different energy spectra and $I_1 \neq I_2$ in thin $WTe_2$ (Fig. 2g) which results from different Fermi velocity of the two edge sides, denoted by the red and blue lines as shown in Fig. 2e. Therefore, the $I(\Phi, \varphi)$ is not symmetric for both $\varphi$ and $\Phi$ anymore. Other possibilities such as vortex trapping, vortex motion during the magnetic field sweep or bulk states asymmetry may contribute. However, the $|I_c^+(B)| = |I_c^-(B)|$ in the thick sample in Fig. 2d helps to largely exclude the other possibility.

We note that two recent preprints[29, 30] has also studied WTe$_2$ Josephson junction and shown evidence of edge states which is explained to be Hinge states[10]. Indeed, it is challenging to unambiguously determine the definite origin of edge superconductivity in our results, and various possibilities exist. However, from the consistent observations of high edge supercurrent density[29], the edge superconductivity is confirmed in the multi-layer WTe$_2$ system. Compared to the reported data, we further perform thickness-dependence experiments and provide more evidence that edge superconductivity exists in thin WTe$_2$ but not a thick one.

We summarize the supercurrent density amplitude ratio of the edge and bulk in Fig. 4. The edge superconductivity gradually emerges in thinner ones and the edge/bulk supercurrent amplitude reaches 2.76 in 10 nm-thick WTe$_2$. The critical thickness for the transition from edge to bulk-dominated superconductivity is estimated to be $t_c = 16 - 20$ nm. Moreover, various topological semimetals such as the TaAs family[31] (Fermi-arc surface states), layered MoTe$_2$[32] (edge states in the 2D limit) and ZrSiS[33] (nodal-line surface states) can be further fabricated into Josephson junctions to detect the surface/edge states.

**Conclusion**

In summary, by studying the Fraunhofer interference, our measurements provide the supercurrent distribution in type-II Weyl semimetal WTe$_2$. In thin WTe$_2$, the existence of edge superconductivity is evidenced. Besides, non-symmetric behavior $I_c^+(B) \neq I_c^-(B)$ in WTe$_2$ through the edge is an intrinsic property of the inversion symmetry breaking, which is distinct from other systems by an external in-plane magnetic field[22]. Furthermore, the Josephson junctions formed from 1D edge states or 2D surface states and $s$-wave superconducting contacts are expected to emulate spinless $p$-wave superconductivity[34] and Majorana flat bands[35] via $a.c.$ Josephson effect by Shapiro response measurements. Edge superconductivity establishes WTe$_2$ as a promising platform for the future realization of topological superconductivity and Majorana bound states.

**Methods**
**WTe$_2$ crystal growth**
High-quality bulk WTe$_2$ crystals were grown by chemical vapor transport (CVT) method as reported before[36]. Single crystals of WTe$_2$ were grown by a high-temperature self-flux method. High-purity tungsten powders (99.9%) and Te pieces (99.999%) were inserted into alumina crucibles with a molar ratio of 1:30 in a glove box filled with pure argon then sealed in quartz tubes under high vacuum. The tubes were heated to 1100 °C in 20 hours and maintained for 10 hours. Then the furnace was slowly cooled down to 650 °C with a rate of 2 °C/h followed by separating the Te flux in a centrifuge at 650 °C.

**Device fabrication**
The WTe$_2$ flakes were mechanically exfoliated onto a Si substrate capped with a 280 nm-thick SiO$_2$ layer and the thickness of WTe$_2$ was identified by optical contrast and atomic force microscopy. The WTe$_2$ Josephson junctions were fabricated by an $e$-beam lithography technique and wet-etched by standard buffered HF solution for 5 s in the electrode regime. We deposited 120 nm-thick Nb electrodes using magnetic sputtering. Then, 40 nm-thick $SiO_2$ was deposited on top to prevent the $WTe_2$ oxidization.

**Transport measurements**

Four-terminal temperature-dependent transport measurements were carried out in a Physical Property Measurement System (PPMS, Quantum Design) with a dilution refrigerator, which achieves a base temperature of 35 mK. The transport properties were acquired using lock-in amplifiers (SR830) and Agilent 2912 meters. We used an excitation current of <50 nA. In differential resistance ($dV/dI$) measurements, a small *a.c.* current bias (10 nA to 100 nA) is generated by the lock-in amplifier output voltage in combination with a 10 M$\Omega$ bias resistor. This small *a.c.* current is added on top of the larger d.*c.* current bias by Agilent 2912, and the induced differential voltage is measured using the lock-in technique with a low frequency (<50 Hz).

**Data availability**

The data that support the plots within this paper and other findings of this study are available from the corresponding author upon reasonable request.

**Acknowledgments**

This work was supported by the National Key Research and Development Program of China (Grant No. 2017YFA0303302 and 2018YFA0305601), the Science and Technology Commission of Shanghai (Grant No. 19511120500), and the National Natural Science Foundation of China (Grant No. 61322407, 11934005, 11874116, 61674040). Part of the sample fabrication was performed at Fudan Nano-fabrication Laboratory. We thank Jinhui Shen from Professor Xiaofeng Jin's group for help in the Nb metal deposition. We thank Liyang Qiu from Saijun Wu's group for help in Matlab code assistance. We thank Quansheng Wu for useful correspondence. AN acknowledges support from Indian Institute of Science. E. Z. acknowledges support from China Postdoctoral Innovative Talents Support Program. We thank Prof. Kam Tuen Law from The Hong Kong University of Science and Technology for helpful discussion on the asymmetric behavior.

**Author contributions**

F.X. conceived the ideas and supervised the overall research. Y.S. and C.Y. synthesized high-quality $WTe_2$ bulk samples. C.H. and E.Z. fabricated the nanodevices. C.H., L. A. and S.L. carried out the PPMS measurements. A.N. and S.S. provided theoretical support. X. X. provided the curve fitting. C.H. and F.X. wrote the paper with assistance from all other authors.

**Competing financial interests**

The authors declare no competing financial interests.

**Supplementary Data**

Supplementary data are available at NSR online.

**Figures and Captions**

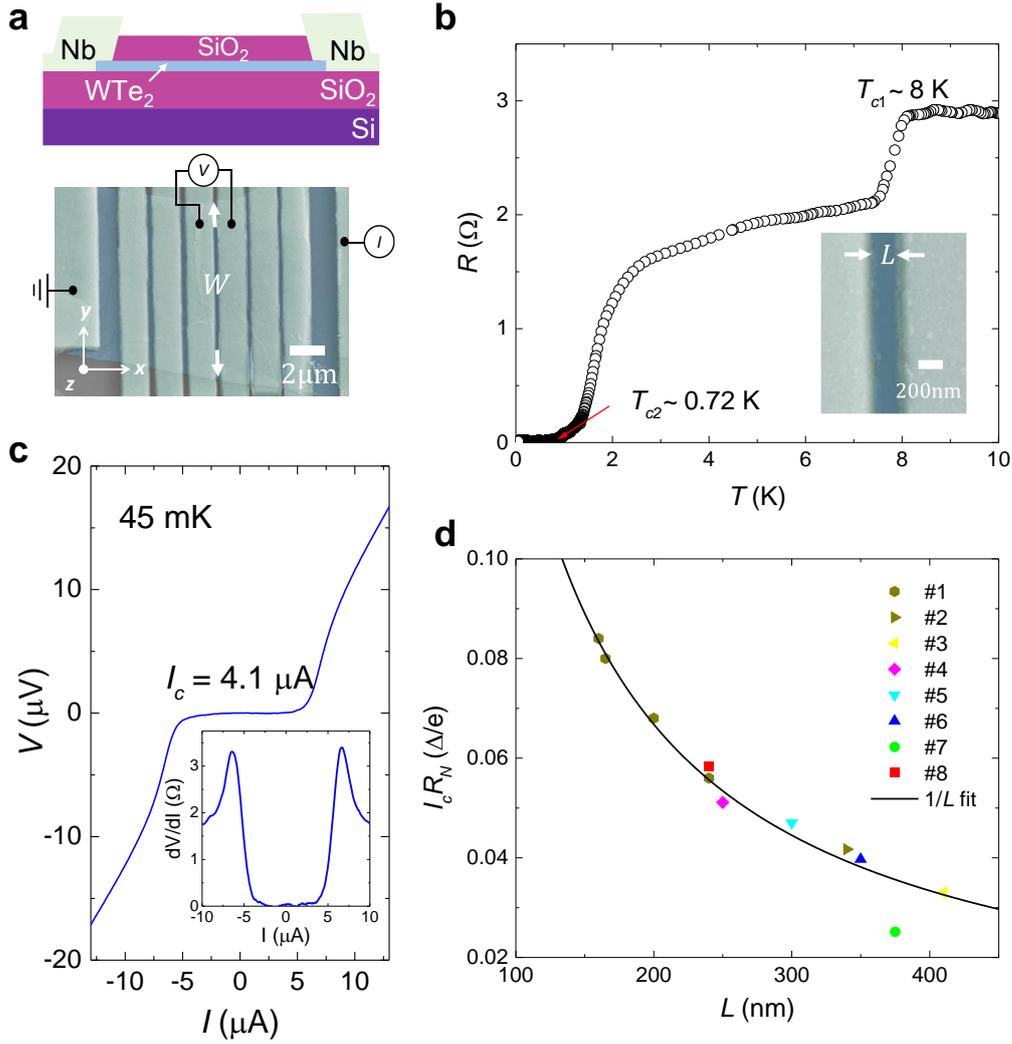

**Figure 1. Josephson effect in thin WTe₂.** (a) Top: Junctions schematic. Bottom: False-colour SEM image of the device with the measurement configuration. 120 nm-thick Nb is deposited on the top of WTe₂ (device #1). A four-terminal measurement across the interface was performed. The in-plane crystal axis of the WTe₂ flake is unknown. (b) Temperature dependence of WTe₂ Josephson junction resistance. Two transitions are identified: $T_{c1} = 8$ K is from the superconducting Nb, $T_{c2} \sim 0.72$ K is from the proximity Josephson coupling of WTe₂. Inset shows that the junction has a length of $L = 200$ nm. (c) *I-V* characteristics for Josephson junction in the superconducting states with a critical current of $I_c \sim 4.1$ μA under zero magnetic field at 45 mK. Inset: d*V*/d*I* characteristics indicate zero resistance below the critical current, the same as the *I-V* curve. (d) Effect of the junction length on supercurrent for eight devices. The product $I_c R_N$ follows a general trend of $I_c R_N \propto 1/L$.

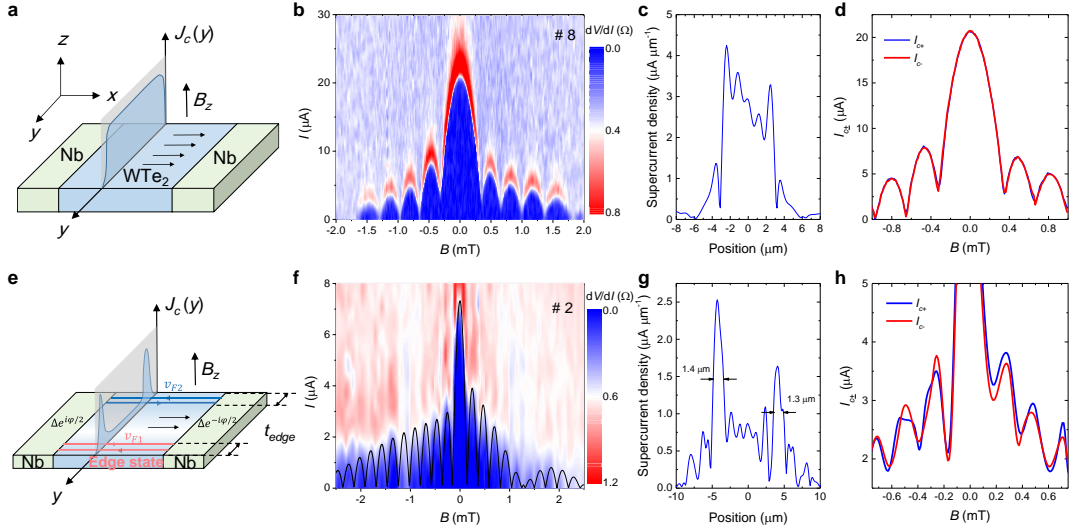

**Figure 2. Evolution of edge superconductivity in thin WTe₂.** (**a**) A schematic picture of a lateral Josephson junction with the out-of-plane magnetic field in thick WTe₂. The thick WTe₂ is filled with charge carriers and the supercurrent can flow uniformly across the junction along the $y$-axis, corresponding to a flat supercurrent density $J_c(y)$. (**b**) The differential resistance at different values of $B_z$ in 60 nm-thick WTe₂ (device #8), showing the single-slit interference characteristics with a uniform supercurrent density. (**c**) The supercurrent distribution along the $y$-axis, which is calculated by the inverse Fourier transform of the data in (**b**). The supercurrent density is uniform along the $y$-axis, consistent with trivial bulk charge transport. (**d**) Critical current $I_c$ as a function of $B$ for the two sweep directions (positive as the blue line, negative as the red line). Two curves overlap with each other. (**e**) A schematic picture of a lateral Josephson junction with the out-of-plane magnetic field where $\Delta e^{\pm i\varphi/2}$ denotes the pairing order parameter of two superconducting Nb electrodes. In thin WTe₂, the bulk domination decreases and the supercurrent is carried by the ~~helical~~ edge ~~states~~. The edges on two sides of WTe₂ have different Fermi velocities $v_{F1}$ and $v_{F2}$ when the inversion symmetry is broken that gives rise to the asymmetric Josephson effect. (**f**) Differential resistance across the 13 nm-thick WTe₂ junction (device #2), showing a mixture of Fraunhofer and SQUID-like pattern with a central lobe of width $< 2\Phi_0$ and side lobes of width $\Phi_0$. The black line shows the fitting results from the edge-stepped supercurrent model. (**g**) The supercurrent distribution of device #2. The widths of the supercurrent-carrying edge channels are estimated to be $1.3 - 1.4\ \mu m$. (**h**) Critical current $I_c$ as a function of $B$ for the two sweep directions (positive as the blue line, negative as the red line), indicating non-symmetric behavior $I_c^+(B) \neq I_c^-(B)$.

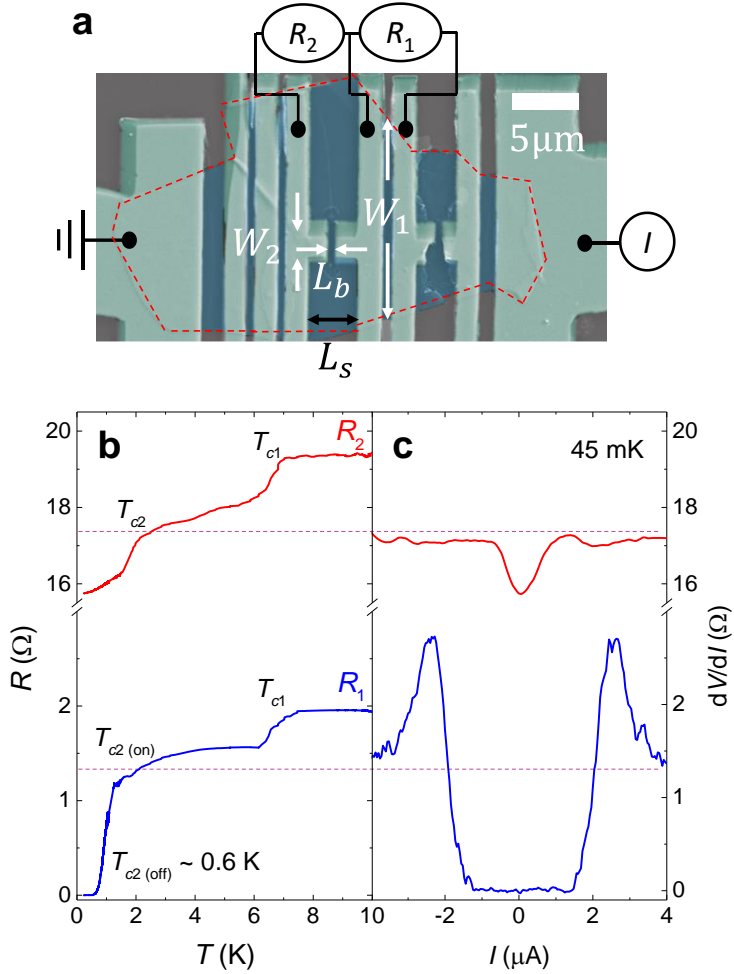

**Figure 3. The coupling of superconductor Nb to the bulk and edge WTe₂ channels. (a)** False-color SEM image of the device #3 with two measurement configurations. $R_1$ and $R_2$ represent the edge-crossing and edge-untouched junctions, respectively. The electrode separation width of $R_2$ from the edge $W_0$ is larger than 5 μm. The length for edge channel $L_s$ is 4 μm while for the bulk channel $L_b \sim 0.4$ μm which makes the edge superconductivity hard to realize. **(b)** Temperature dependence of resistance in two junctions as shown in (a). $T_{c1}$ is the superconducting transition of Nb while the superconducting WTe₂ emerges at $T_{c2}$. **(c)** d$V$/d$I$ characteristic at 45 mK.

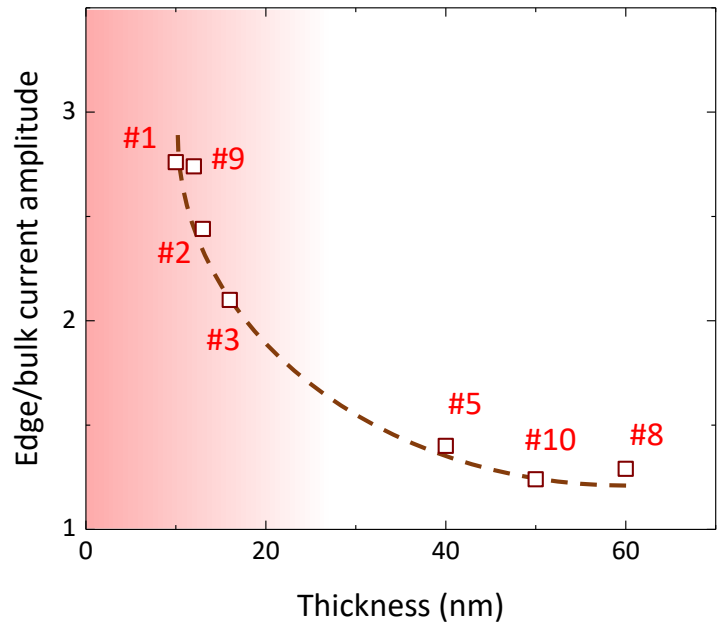

**Figure 4. Summary of thickness-dependent edge-supercurrent-density contribution in WTe$_2$.** The edge/bulk current amplitude is estimated by the ratio of edge/bulk ($\frac{j_{edge}}{j_{bulk}}$) supercurrent density. $j_{edge}$ and $j_{bulk}$ are estimated by the average value of the left and right peaks and the value of the central region in position-dependent supercurrent density. The dashed line shows the trend.